\documentclass[prx, twocolumn, times, longbibliography, aps, amssymb, footinbib, showpacs, superscriptaddress]{revtex4-1}
\usepackage{amsmath, braket, dsfont, units}
\usepackage{graphicx}				
\usepackage{hyperref}

\usepackage[dvipsnames]{xcolor}
\definecolor{Zcolour}{rgb}{0.992, 0.588, 0.22}
\definecolor{dkgreen}{rgb}{0,0.5,0}
\definecolor{purple}{rgb}{0.5,0,0.5}

\newcommand{\footnoteremember}[2]{
\footnote{#2}
\newcounter{#1}
\setcounter{#1}{\value{footnote}}
}
\newcommand{\footnoterecall}[1]{\footnotemark[\value{#1}]}

\newcommand{\ldos}{\mathrm{LDOS}}
\newcommand{\di}{d_\mathrm{i}}
\newcommand{\dg}{d_\mathrm{g}}

\begin{document}
\title{Imaging anyons with scanning tunneling microscopy}
\author{Zlatko Papi\'c}
\affiliation{School of Physics and Astronomy, University of Leeds, Leeds LS2 9JT, United Kingdom}
\author{Roger S. K. Mong}
\affiliation{Department of Physics and Astronomy, University of Pittsburgh, Pittsburgh, PA 15260, United States}
\author{Ali Yazdani}
\email{yazdani@princeton.edu}
\affiliation{Department of Physics, Princeton University, Princeton, NJ 08540, U.S.A.}
\author{Michael P. Zaletel}
\email{mzaletel@princeton.edu}
\affiliation{Department of Physics, Princeton University, Princeton, NJ 08540, U.S.A.}

\begin{abstract}
Anyons are exotic quasi-particles with fractional charge that can emerge as fundamental excitations of strongly interacting topological quantum phases of matter.
Unlike ordinary fermions and bosons, they may obey non-abelian statistics--a property that would help realize fault tolerant quantum computation.
Non-abelian anyons have long been predicted to occur in the fractional quantum Hall (FQH) phases that form in two-dimensional electron gases (2DEG) in the presence of a large magnetic field, such as the $\nu=\tfrac{5}{2}$ FQH state.
However, direct experimental evidence of anyons and tests that can distinguish between abelian and non-abelian quantum ground states with such excitations have remained elusive.
Here we propose a new experimental approach to directly visualize the structure of interacting electronic states of FQH states with the scanning tunneling microscope (STM).
Our theoretical calculations show how spectroscopy mapping with the STM near individual impurity defects can be used to image fractional statistics in FQH states, identifying unique signatures in such measurements that can distinguish different proposed ground states.
The presence of locally trapped anyons should leave distinct signatures in STM spectroscopic maps, and enables a new approach to directly detect - and perhaps ultimately manipulate - these exotic quasi-particles.
\end{abstract}

\maketitle

	Unlike spontaneous symmetry breaking, which is characterized by a local order parameter, detecting topological order requires a ``non-local'' probe sensitive to the fractional phases anyons accrue when they wind around each other \cite{Arovas1984}.
Detecting these phases would seem to require the  construction of an ``anyon interferometer'' in which one test particle is controllably brought around another,  as has been proposed for FQH systems \cite{Chamon1997, Bonderson2008, Willett2013}.
But in fact spectroscopy - even an atomically local one like STM -  is inherently non-local:
accurately resolving  energies detects properties of quasiparticles separated by long times. 
Previous proposals for topological spectroscopy include the tunneling conductance of FQH edge states,\cite{ChangRMP}  the threshold behavior of  neutron scattering in spin-liquids,\cite{Morampudi} and the structure of RF absorption lines of trapped cold-atomic FQH  droplets \cite{CooperSimon}.
While these experiments could narrow down the nature of the topological order, they could not be used to image the presence of individual anyons localized in the bulk.

Until recently, STM in the bulk of a FQH state was largely impractical, since the pristine 2DEGs required were only possible to realize in quantum wells sandwiched deep between semiconductors (though workarounds have been proposed \cite{DGG1, DGG2}).
The observation of QH effects in graphene \cite{Zhang2005} and the surface states of bismuth,\cite{hofmann2006surfaces} where the 2DEG is atomically close to the vacuum, removes this  obstacle.
In these systems, STM has not only been able to detect single particle quantization, which gives rise to the integer quantum Hall effect (IQHE), but has also been able to detect interaction induced exchange gaps that results in quantum Hall ferromagnetism \cite{Stroscio2010high, Stroscio2010real, Luican2014, Feldman2016}.
More relevant to our studies here, recent experiments have found that when the samples contain very few defects (a single defect in a magnetic length), STM can not only directly probe the spatial structure of the Landau orbits but also use this capability to visualize when the ground state breaks the symmetry of the underlying lattice \cite{Feldman2016}.
These advances together with the now robust observation FQHEs in graphene, where potential non-Abelian states are observed, \cite{Morpurgo2014, Zibrov2016, Dean2017} makes examining the prospects for STM studies in the fractional regime a timely topic. 

What could STM spectroscopy and its ability to make spatially resolved measurement reveal about the nature of topological excitations in the FQH?
Assuming that STM as function of $E=eV$ at a location $\mathbf{r}$ measures the local density of states (LDOS), i.e., the differential conductance $\frac{dI}{dV}(\mathbf{r}) \propto \ldos(e V, \mathbf{r})$, then in the bulk of a perfectly clean insulating QH system such measurements should simply show a charging gap and would be spatially featureless.
Approaching the edges the behavior of the the spectra would change, where in contrast to the bulk, tunneling spectra should show power-law energy dependence $E^g$\cite{ChangRMP}.
So naively, besides potential for edge spectroscopy one might expect that local STM measurements would not be a useful tool to study anyons in the bulk of a FQH phase. 

The key insight, however, is to consider the FQH phases when there is a single isolated impurity, (e.g., a lattice defect or a pinned charge) and examine the LDOS on the length scale of the magnetic length surrounding this impurity.
As we demonstrate here, the LDOS develops spatial modulations near the impurity due to the rich spectrum of discrete bound states.
An electron injected by the STM splits into $q$ charge $-e/q$ anyons, so the local STM spectrum should be interpreted as the distinct energies at which  $q$ anyons can be bound to the impurity.
Remarkably, we find that the counting of the impurity bounded levels is sensitive to ``fractional exclusion statistics'' of anyons, which as first considered by Haldane \cite{Haldane1991, MurthyShankar1994, CooperSimon} interpolates between statistics of bosons and fermions.
Using both model wavefunction calculations and numerical exact  diagonalization studies, we find that the local spectrum can be a powerful new method to detect underlying topological order in the system, as different abelian and non-abelian phases have their own distinct spectrum near the impurity.
More importantly, whether the impurity binds an anyon or not also modifies the local spectrum, thereby making it possible to use the STM as a true ``anyon microscope''  which can detect anyons regardless of their electrical charge (see for example, \cite{venkatachalam2010local} for charge sensitive approach).
We explore how these ideas many be implemented in experiments on graphene by considering realistic Coulomb interactions and impurities relevant to this system.

\begin{figure}[t]
\includegraphics[width=0.9\columnwidth]{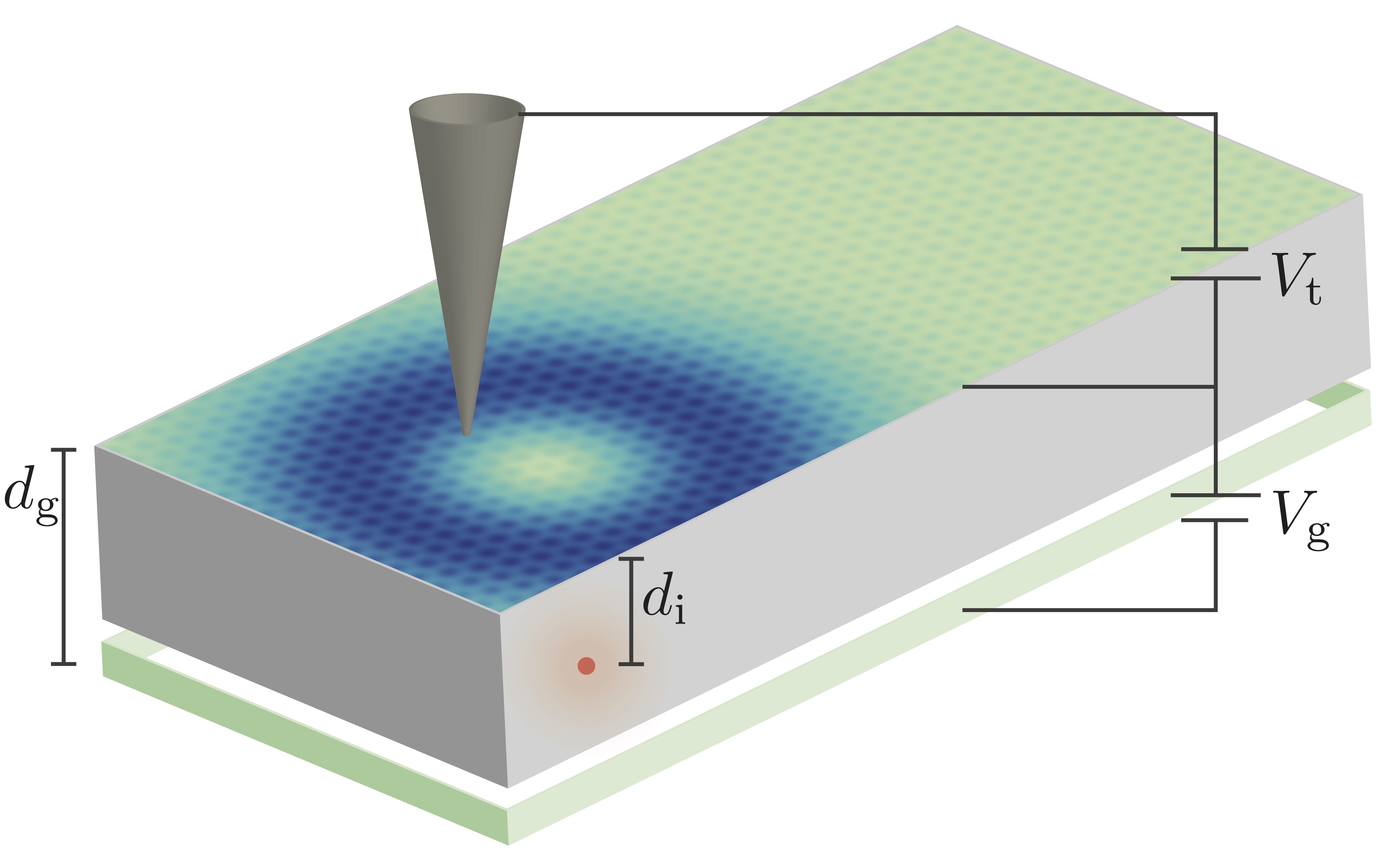}
\caption{\textbf{Experimental setup}. A metallic gate (bottom) is separated by an insulating barrier of thickness $\dg$ from a 2DEG (top). An impurity of charge $Z$ is located inside the barrier at distance $\di$ from the 2DEG.
Tuning the gate voltage $V_\textrm{g}$ to enter into a quantum Hall state, STM spectroscopy is used to measure the local density of states (LDOS), shown as a density plot.
The LDOS will reveal a discrete set of ring-like resonances centered on the impurity, whose radius depends on the energy.
By counting the number of resonances of a given radius, it is argued that the location and fractional exclusion statistics of the anyons can be determined.}
\label{fig:sandwich}
\end{figure}

\emph{Hamiltonian and LDOS.}
We begin with a description of the FQH setup illustrated in Fig.~\ref{fig:sandwich}.
We consider a 2DEG at a distance $\dg$ above a metallic gate in a quantizing magnetic field $B$, as realized for example in a graphite/boron nitride/graphene heterostructure.
Due to  the image charges induced by the gate~\cite{Papic11,Hu11}, the Coulomb interaction between electrons in the 2DEG is screened and takes the form 
\begin{equation}\label{eq:vr}
	V_C(r) = \frac{1}{r} - \frac{1}{\sqrt{(2 \dg)^2 + r^2}},
\end{equation}
which falls off as $r^{-3}$ for $r \gg \dg$. 
In addition, we assume an impurity  with charge $Z$ is located a distance $\di < \dg$ below the origin of the 2DEG, leading to a one-body potential  
\begin{equation}\label{eq:ur}
	U(r) = \frac{Z}{\sqrt{\di^2 + r^2}} - \frac{Z}{\sqrt{(2 \dg - \di)^2 + r^2}}.
\end{equation}
In the case of graphene devices, such impurities can arise from substitutions in the boron nitride \cite{Crommie2014}.
Below we set $Z=\pm 1$ and treat $\di$ as a tunable parameter. 

In a strong perpendicular magnetic field $B$, the kinetic energy is quenched and topological phases, such as Laughlin~\cite{Laughlin:1983} and composite fermion states~\cite{Jain89}, emerge as ground states of $V_C$ at particular filling fractions $\nu=N/N_\phi$, where $N$ is the number of electrons and $N_\phi$ is the magnetic flux piercing the 2DEG.
The states which appear are insensitive to $\dg$ once $\dg \gg \ell_B$, where  $\ell_B = \sqrt{\hbar/eB}$ is the magnetic length, so below we fix $\dg = 4\ell_B$. 

We begin by ignoring any influence of the tip on the measurement, where STM  $\frac{dI}{dV}$  spectra at positive tip bias relative the sample is proportional to the occupied LDOS,
\begin{align}
	\ldos(E = e V, \mathbf{r}) & =  \sum_a \delta(E - E_a)   |\bra{a} \hat{\psi}(\mathbf{r}) \ket{\Omega}|^2.
	\label{eq:ldos}
\end{align}
Here $\ket{\Omega}$ is the ground state, $\hat{\psi}(\mathbf{r})$  is the electron destruction operator, and $\ket{a}$ runs over all excited states with one less electron than $\Omega$.
The energy $E_a$ of the excited state is measured relative to the difference in work function between the ground state of the sample and the tip, a point we will return to.
An analogous expression holds for the unoccupied LDOS with $\hat{\psi} \to \hat{\psi}^\dagger$.
Because quantum Hall states are gapped, a finite bias is required before any tunneling occurs.

\begin{figure*}[t]
	(a) LDOS for $\nu=1$
	\includegraphics[width=0.98\textwidth]{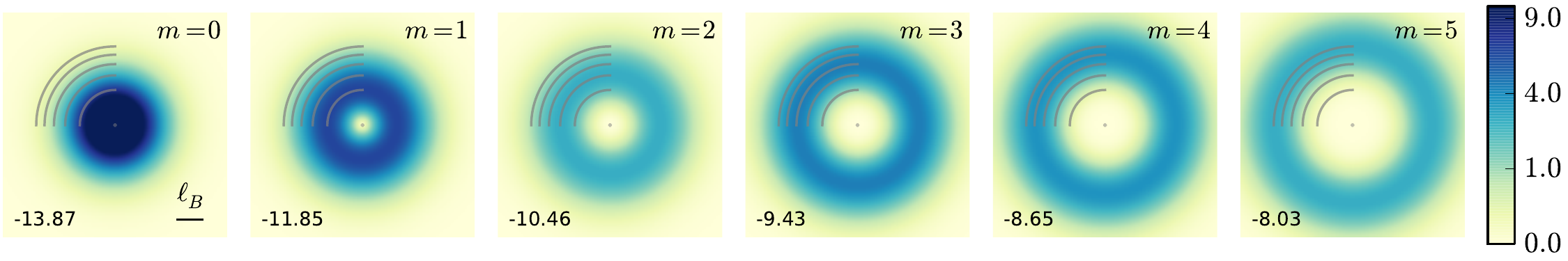}
\\	(b) LDOS for $\nu=\frac{1}{3}$
	\includegraphics[width=0.98\textwidth]{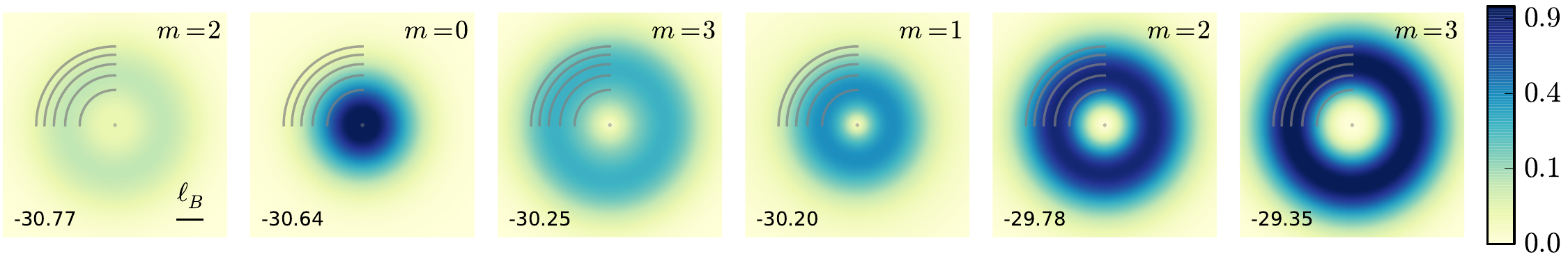}
	\caption{%
		\textbf{The simulated occupied} $\ldos(E, \mathbf{r} = x, y)$ \textbf{for the} $\nu = 1$ \textbf{IQHE and} $\nu = \frac{1}{3}$ \textbf{Laughlin phase}. The calculations assume a dipolar Coulomb interaction (equation \eqref{eq:vr}) and charged impurity located below the origin (equation \eqref{eq:ur}).
		Each image is at a fixed $E$ annotated in the lower left in units of \unit{meV} at $B=\unit[14]{T}$, and a scale bar indicates $\ell_B \sim \unit[7]{nm}$. Away from the displayed $E$, there is no DOS.
		The white arcs denote the density maxima $r_m$ of the $m$\textsuperscript{th} single particle orbital.
		(a) For $\nu = 1$, each orbital lights up once while sweeping through the energy, indicating there is a single hole-excitation for each angular momentum $m$.
		(b) For $\nu = \frac{1}{3}$, the $m=2$ ring lights up \emph{twice}, once at energy $\unit[-30.77]{meV}$, and once again at $\unit[-29.78]{meV}$.
		This indicates there are multiple many-body excitations associated with removing an electron from orbital $m$, as a result of strong correlations.
		The parameters $\dg, \di$ are given in Fig.~\ref{fig:LDOS}.
}
	\label{fig:fhqe_rs}
\end{figure*}

The Hamiltonian is strongly interacting, so to compute the LDOS exactly we must resort to numerical diagonalization of a finite number of electrons on a sphere, where the bulk properties can be conveniently probed without edge effects~\cite{Haldane83, Patton2014} (see Methods).
In Fig.~\ref{fig:fhqe_rs}, we show the resulting prediction for the occupied LDOS  near the impurity, both for the IQHE ($\nu = 1$) and the Laughlin phase ($\nu = \frac{1}{3}$).
The LDOS reveals a discrete set of ring-like features centered around the impurity whose radius and intensity depends on the energy.

To understand the ring features in the LDOS, we turn to the single-particle physics of the LL.
In strong magnetic fields, the many-electron system can be modeled in a restricted Hilbert space of the given Landau level (LL)~\cite{PrangeGirvin}.
In the lowest LL relevant to fillings $\nu = \frac{1}{3}$ and $1$, the single-electron states on the plane (in the symmetric gauge) are given by
\begin{equation}\label{eq:phim}
	\phi_m(\mathbf{r}) \propto r^{m} e^{i\phi m} e^{-r^2/(4\ell_B^2)}, \;\;\; m = 0, 1, 2, \dots, 
\end{equation}
where $(r, \phi)$ are polar coordinates.
Notably, their density, $|\phi_m(\mathbf{r})|^2$, is concentrated in rings of radii $r_m = \sqrt{2 m} \ell_B$ around the impurity.
To good approximation, the Hamiltonian of the 2DEG then consists of potentials $V(r)$ and $U(r)$ projected into the many-body Hilbert space of the lowest LL.  

To compute the LDOS at energies small compared to the cyclotron gap between LLs, the electron operator can be expanded as $\hat{\psi}(\mathbf{r}) = \sum_m \phi_m(\mathbf{r}) \hat{c}_m$, where $\hat{c}_m$ removes an electron from orbital $m$.
If we further assume that the impurity potential is rotationally symmetric on the scale of $\ell_B$, so that the angular momentum $L^z = \hbar m$ is a good quantum number, we can insert this expansion into equation \eqref{eq:ldos} to obtain
\begin{align}
\ldos(E, \mathbf{r}) & =  \sum_{a} \delta(E - E_a)  |\phi_{m_a}(r)|^2 |\bra{a} c_{m_a} \ket{\Omega}|^2
\label{eq:ldosm}
\end{align}
where $\hbar m_a$ is the angular momentum of excited state $a$.
The rings that appear in the LDOS thus arise because each excited state $a$ only contributes to the LDOS around a ring $|\phi_{m_a}(r)|^2$ of radius $r_{m_a}$. 

\begin{figure*}[t!]
	\includegraphics[width=0.98\textwidth]{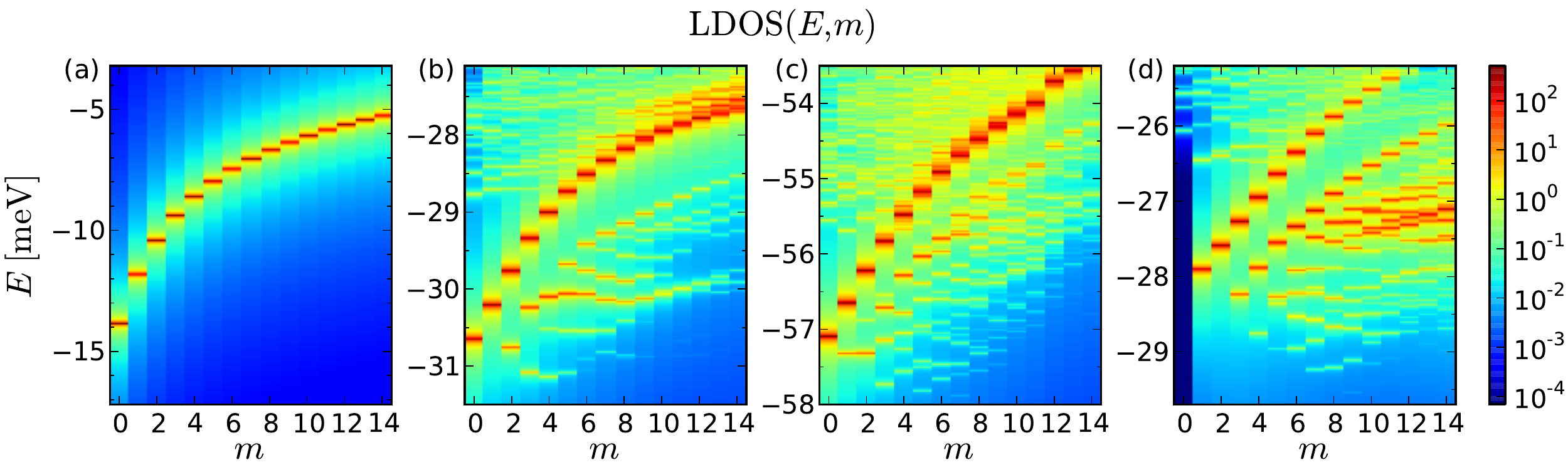}    
	\caption{\textbf{The orbital-resolved $\ldos(E, m)$ for various states.}
	In all cases a phenomenological width of \unit[7]{$\mu$eV} was added to the data.
	(a) The IQH state. There is one level for each $m$.
		The impurity has charge $e$ and is situated at $d_i =3.1\ell_B$ below the 2DEG.
	(b)  The $\nu = \frac{1}{3}$ FQH state, with no anyons localized on the impurity.	
	There  are now multiple energy levels for each $m$. The counting of the low lying levels (i.e., the number of excitations at each $m$) is $1, 1, 2, 3, 4, 5, \ldots$. States above the brightest band form a non-universal continuum. The system contains $N_\phi=24$ flux quanta with impurity at $d_i=2.7\ell_B$.
	(c) The $\nu = \frac{5}{2}$ Moore-Read state. The counting of the low lying levels is $1, 2, 3, 5,\ldots$. In order to break particle-hole symmetry, a small component of three-body interaction $V_{3b} = 0.1$ was added to the screened Coulomb Hamiltonian. The system contains $N_\phi=25$ flux quanta with impurity at $d_i=2.3\ell_B$.
 	(d) Same as (b) but in the presence of a charge-$\frac{e}{3}$ anyon bound to the impurity. The counting of the low lying levels is now $0, 1, 1, 2, 3, \ldots$. The system contains $N_\phi=24$ flux quanta with impurity at $d_i=2.8\ell_B$.
	}
	\label{fig:LDOS}
\end{figure*}

The ring structures in the LDOS at high magnetic field allows us to convert the real-space information into an angular momentum $m$,\footnote{Indeed, guiding-center dynamics imply that the quadrupole moment and angular momentum of an excitation are are related by $Q_{zz} = e \ell_B^2 L^z / \hbar$.}
so STM can be thought of as $L^z$ resolved (at least for low $m$ - the inverse problem becomes ill-conditioned for large $m$, where the $r_m = \sqrt{2 m} \ell_B$  become too closely space compared with $\ell_B$ to accurately assign).
Thus rather than plotting the $\ldos$ in real space, it is  more compact to plot the intensity in the $m$\textsuperscript{th} ring, 
\begin{align}
	\ldos(E, m) & =  \sum_a \delta(E - E_a) \, \big|\!\braket{a | \hat{c}_m | \Omega}\!\big|^2
\label{eq:mldos}
\end{align}
as shown in Fig.~\ref{fig:LDOS}. It reveals a set of energies that disperse with $m$.

\emph{IQHE.} This dispersion is easiest to understand at $\nu = 1$,\cite{Prange81} shown in Fig.~\ref{fig:LDOS}(a). 
The ground state $\ket{\Omega}$ is a full LL, while the excited states $a$ consist of a hole in orbital $m = a$, $\ket{a} = c_a \ket{\Omega}$.
The impurity potential decomposes into the energies for occupying each orbital: $\hat U = \sum_{m=0}^\infty U_m \hat{n}_m$, where $U_m  \sim U(r_m)$ characterizes the radial fall-off of the impurity potential. 
Thus the tunneling DOS into orbital $m$ is shifted down in energy by $U_m$ relative to the region far from the impurity.
There is one state per $m$, i.e., the ``counting'' is $1, 1, 1, \dots$
(inter-LL excitations, which we ignore, occur at a much higher energy).

STM experiments have already shown that such impurity-induced shifts of the orbital levels near single impurities can be resolved experimentally in the IQHE case.
In the case of a graphene on a boron-nitride heterostructure, spectroscopic measurements near individual impurity could resolve levels corresponding to $m = 0, 1, 2$ orbitals of the $N=0$ LL, while higher $m$ appeared as a continuum \cite{Luican2014}.
In studies of $\mathrm{Bi}$ surface states at high magnetic fields~\cite{Feldman2016}, atomic sized impurities were found to shift the $L^z = 0$ orbital in each of the $N$\textsuperscript{th} LL (for both electron and hole like LL), and could not only be detected in energy resolved experiments but also spatially mapped with high resolution.
The latter experiments demonstrate that in samples with low defect density (less than one per magnetic length) STM spectroscopic maps are capable of resolving features such as those in Fig.~\ref{fig:LDOS}(a). 

\emph{Fractional phases.}
We now examine how the structure of the LDOS near impurities differs between integer and fractional cases.
The $\nu = \frac{1}{3}$ case, Fig.~\ref{fig:LDOS}(b) demonstrates this difference by showing that rather than one state per $m$, most of the $m$  have \emph{multiple} energies appearing in the LDOS.
Although some of these states belong to a high-energy continuum, there is also a band of discrete low-energy levels whose number grows with angular momentum $m$  as $1, 1, 2, 3, 4, 5, 7, \dots$. 
This multiplicity is a signature of fractionalization. When an electron is removed from the Laughlin state, the hole fractionalizes into  three independent  charge $e/3$ quasiholes~\cite{Laughlin:1983}.
Because they can have  motion relative to each other,  there are multiple three-quasihole states of a given total angular momentum $\hbar m$.
The quasiholes have a ``topological interaction,"  their fractional statistics, which results in  a $2 \pi/3$ Berry phase whenever one winds around another.
The fractional statistics leave their imprint on the angular momentum, and hence counting, of the three-quasihole states.
Note there is no ``orthogonality catastrophe'' in this situation, since almost the entirety of the spectral weight lies in this discrete set of low energy states: while the lower-energy charge $e/3$ excitation is orthogonal to the electron, the electron itself is readily reconstructed from three of them \cite{Jain2005}.

\begin{table}[htb]
	\newcommand{\z}{$\cdot$}\newcommand{\X}{$\bullet$}
	\begin{minipage}[t]{40mm} Laughlin\\
	\begin{tabular}[t]{c|cccccccccc}
		$m$ & 0 & 1 & 2 & 3 & 4 & 5 & 6 & 7 & 8 & 9
	\\\hline\hline
	$\ket{\Omega}$ & \X & \z & \z & \X & \z & \z & \X & \z & \z & \X
	\\\hline
	0  & \z & \z & \z & \X & \z & \z & \X & \z & \z & \X
	\\\hline
	1  & \z & \z & \X & \z & \z & \z & \X & \z & \z & \X
	\\\hline
	2  & \z & \X & \z & \z & \z & \z & \X & \z & \z & \X
	\\ & \z & \z & \X & \z & \z & \X & \z & \z & \z & \X
	\\\hline     
	3  & \X & \z & \z & \z & \z & \z & \X & \z & \z & \X
   \\ & \z & \X & \z & \z & \z & \X & \z & \z & \z & \X
   \\ & \z & \z & \X & \z & \z & \X & \z & \z & \X & \z
	\\\hline\multicolumn{11}{c}{Laughlin + $e/3$ hole}\\\hline
	$\ket{\frac{e}{3}}$ & \z & \X & \z & \z & \X & \z & \z & \X & \z & \z \\
	\hline 
	1 & \z & \z & \z & \z & \X & \z & \z & \X & \z & \z \\
	\hline 
	2 & \z & \z & \z & \X & \z & \z & \z & \X & \z & \z \\
	\hline
	\end{tabular} \end{minipage}
	\begin{minipage}[t]{45mm} Moore-Read\\
	\begin{tabular}[t]{c|ccccccccccccc}
		$m$ & 0 & 1 & 2 & 3 & 4 & 5 & 6 & 7 & 8 & 9 & \multicolumn{2}{c}{$\cdots$}
	\\\hline\hline
	$\ket{\Omega}$ & \X & \X & \z & \z & \X & \X & \z & \z & \X & \X & \z & \z & \X
	\\\hline
	0  & \z & \X & \z & \z & \X & \X & \z & \z & \X & \X & \z & \z & \X
	\\\hline
	1  & \X & \z & \z & \z & \X & \X & \z & \z & \X & \X & \z & \z & \X
	\\ & \z & \X & \z & \X & \z & \X & \z & \z & \X & \X & \z & \z & \X
	\\\hline
	2  & \X & \z & \z & \X & \z & \X & \z & \z & \X & \X & \z & \z & \X
	\\ & \z & \X & \X & \z & \z & \X & \z & \z & \X & \X & \z & \z & \X
	\\ & \z & \X & \z & \X & \z & \X & \z & \X & \z & \X & \z & \z & \X
	\\\hline
	3  & \X & \z & \X & \z & \z & \X & \z & \z & \X & \X & \z & \z & \X
	\\ & \X & \z & \z & \X & \X & \z & \z & \z & \X & \X & \z & \z & \X
	\\ & \X & \z & \z & \X & \z & \X & \z & \X & \z & \X & \z & \z & \X
	\\ & \z & \X & \X & \z & \z & \X & \z & \X & \z & \X & \z & \z & \X
	\\ & \z & \X & \z & \X & \z & \X & \z & \X & \z & \X & \z & \X & \z
	\\\hline
	\end{tabular} \end{minipage}
	
	\caption{%
	\textbf{Computing the LDOS counting of the Laughlin and Moore-Read state.}
	The low-energy states of many FQH phases can be captured by their ``root configuration,'' which serves as a representative cartoon of the many-body wavefunction.
	Each root configuration represents a state $\ket{n_0, n_1, \cdots}$ specifying the electron occupation $n_m$ of orbital $m$.
	In this table, every row is a root configuration with the empty/filled orbitals  denoted by \z/\X\ symbols.
	The low-energy root configurations of a FQH phase satisfy a ``$(k,r)$ exclusion rule''.
	The $\nu = \tfrac{1}{3}$ Laughlin phase satisfies the $(1,3)$-rule: there is at most 1 particle in every 3 consecutive orbitals.
	The Moore-Read phase satisfies the $(2,4)$-rule: no more than 2 particles within 4 neighboring orbitals.
	In each case, the ground state $\ket{\Omega}$ corresponds to the densest configuration of particles obeying the exclusion rule.
	The low-energy charge $e$ states are enumerated by root configurations with one less particle than $\ket{\Omega}$ which \emph{still} obey the $(k,r)$-rule.
	As the angular momentum is $L^z = \hbar \sum_m m \, n_m$, we can compute the angular momentum of the charge $e$ states relative to the ground state, which is denoted in the first column.
	The number of possible configurations of a given angular momentum ($1,1,2,3,etc.$ for the Laughlin and $1,2,3,5,etc.$ for the Moore-Read), is the counting of the LDOS.
	In the case of the $\tfrac{1}{3}$ phase, the root configuration in the presence of a trapped quasihole, $\ket{e/3}$, is obtained by shifting $\Omega$ outwards by one.
	}
	\label{tab:vacuum_count}
\end{table}

	More generally, in a FQH phase where the hole splits into $q$ charge $e/q$ anyons, the orbitally-resolved LDOS reveals all the distinct eigenstates in which the $q$ anyons have  total angular momentum $\hbar m$.
While we do not know how to compute this counting in complete generality, for many FQH states the observed sequence can be  predicted using  ``$(k,r)$ fractional exclusion statistics''~\cite{Haldane1991, BernevigHaldane2008}, as explained in Tab.~\ref{tab:vacuum_count}.
All FQH phases captured by one (or more) $(k,r)$-rule, including the Laughlin and $\mathbb{Z}_k$ Read-Rezayi sequences,  have  distinct LDOS counting sequences. 
For example, the non-Abelian Moore-Read state~\cite{MooreRead1991}, whose LDOS we show in Fig.~\ref{fig:LDOS}(c), is predicted by Tab.~\ref{tab:vacuum_count} to have counting $1, 2, 3, 5,\dots$.
In this sense the impurity LDOS provides a ``fingerprint'' for identifying the underlying topological order. 
Note that this counting is not the same as the counting of the edge excitations, though it is closely related to the counting of the ``particle entanglement spectrum'' which has been used to numerically identify FQH phases \cite{BernevigPES}.

We now discuss some variants of this result.
From a topological perspective, an electron and hole are equivalent.
Nevertheless, the occupied LDOS (which encodes $|\!\braket{a | \hat{c}^\dagger_m | \Omega}\!|^2$) has a different counting than the unoccupied side.
While this counting cannot be computed from the simplest version of the $(k,r)$-rule, for FQH states belonging to the Jain sequence it can be predicted by appealing to the composite fermion picture~\cite{Jain89}. 
For $\nu = \frac{1}{3}$, our numerics agree with the predicted electron counting $0, 0, 0, 1, 1, 2, 3, 4, \dots$ in the lowest band of states, which is  shifted to larger $m$ relative to the hole-side\mbox{\footnoteremember{fn}{See Supplementary Online Material.}\hspace{-1.5ex}.}
The spectral weight of this band is small ($\sim 10^{-3}$) but appears to remain finite in the thermodynamic limit.

The difference between the occupied and unoccupied LDOS could provide a valuable probe of the half-filled Landau level, where it has proved
extremely difficult to experimentally distinguish between  the Pfaffian~\cite{MooreRead1991}, anti-Pfaffian~\cite{Levin07,Lee07} and PH-Pfaffian~\cite{Son2015} phases, as they have identical Hall conductance and shot-noise signatures. 
These states are the leading theoretical candidates for the half-filled  plateaus of $\mathrm{GaAs}$ \cite{Willett87} and bilayer graphene \cite{Morpurgo2014, Zibrov2016, Dean2017}.
At half-filling, the system has an approximate particle-hole (PH) symmetry which relates filled and empty orbitals of the valence LL.
The hole and electron side of the LDOS are related by PH, so any difference between them at half-filling provides a sharp probe of PH-symmetry breaking.
The Pfaffian and anti-Pfaffian states are particle-hole conjugates of each other (and hence break PH), while the PH-Pfaffian is PH-symmetric.
If the counting of the LDOS on the hole and electron side are found to be identical, this is strong evidence for the PH-Pfaffian (whose counting has not been predicted).
On the other hand, the Pfaffian hole counting ($1, 2, 3, 5, \dots $) will differ from its electron counting; while we are unable to calculate the latter in full, earlier calculations \cite{Hansson2009, BernevigHaldane2009, Rodriguez2012} suggest counting $1, 0, \dots$.
The anti-Pfaffian is analogous but with the role of electron and hole reversed.
Thus observing hole counting $1, 2, \dots$ would be strong evidence for the Pfaffian, and visa-versa.

In the calculations thus far, the impurity was weak enough that it did not trap a charge in the ground state.
However, at an electron density slightly away from the center of the plateau, there will be a finite density of quasiparticles pinned by the disorder.
A quasiparticle localized on the impurity will leave its imprint on the allowed quasihole states injected by the STM, since they experience a statistical interaction.
In Fig.~\ref{fig:LDOS}(d) we show the $\nu = \frac{1}{3}$ hole-LDOS in the presence of a trapped $\frac{e}{3}$ quasihole.
We find distinct counting $0, 1, 1, 2, 3, 5, 6, \dots$, as derived from $(k, r)$ statistics in Tab.~\ref{tab:vacuum_count}.
Most strikingly, there is a complete absence of tunneling from the $m=0$ orbital, providing a discrete signature of the fractionalized quasiparticle which can be directly observed in STM.
The LDOS near the trapped quasiparticles of the Pfaffian phase also prove to be distinct.\footnoterecall{fn}
Thus the LDOS can be used to image the anyon charge localized to each impurity.

\emph{Experimental considerations.}
	As described earlier, STM experiments on the IQHE have already observed how impurities can shift individual orbitals within a LL \cite{Luican2014, Feldman2016} and hence they strongly support our proposal for similar STM experiment that can detect anyons and their statistics in the FQH state.
However, the previous experiments and their extensions to the FQH state, for example on high quality graphene devices, still raise several experimental issues.
We identify three potential experimental challenges that need to be addressed: 1) tip-induced band bending, 2) charging and life-time effects, 3) tip-induced screening of the electron-electron interactions that give rise to FQH states. 

	Typically the work function of the STM tip is different than that of the sample, and as a result there in an effective electric field present in the junction beyond that due to the tip-sample voltage bias.
In the case of semiconducting samples, this difference combined with poor dielectric screening results in band-bending underneath the tip.
Ideally, the experiments proposed here on states with poor screening can  be carried out with tips that have a work function matching the sample; for graphene, carbon coated tips, or  perhaps transition metal tips that have been in situ prepared with growth of a graphene monolayer on their surface.
So far there has not been substantial effort in engineering STM tips for this purpose.
However, even without the ideal tips, but with use of a backgate, as we consider in our experimental geometry Fig.~\ref{fig:sandwich}, the local chemical potential underneath the tip can still be adjusted to make the proposed experiments possible.
Successful use of a backgate in graphene devices while performing STM spectroscopy of LL has been previously accomplished \cite{Chae2012}.
These backgate mapping spectroscopic measurements also show how the influence of the tip can be quantified experimentally, thereby providing a method which in combination with tip engineering can be used to minimize the effect of tip-induced band bending. 

	Another natural concern is whether charging effects could complicate the proposed experiments, since we are proposing tunneling into a localized states of a QH insulator.
Previous studies of impurity-shifted LL orbital states in graphene and $\mathrm{Bi}$ \cite{Luican2014, Feldman2016} do not appear to suffer from this problem, mostly likely because they are carried out in the regime where the tunneling rates (with pico or nano amp currents) are smaller than the inverse life-time of the localized states that are probed.
In our proposal, we would like the broadening to be large enough so that charging is not an issue, but not too large to wash out the fine structure in our spectrum.
Since a \unit{nA} tunneling current corresponds to a \unit[100]{ps} lifetime, while the features are split on the order of \unit[10]{ps}, this constraint should be possible to satisfy experimentally.
The constraint can be verified by checking for linear response between the current and tunneling matrix element, as done for Shiba bound states of a superconductor~\cite{Ruby2015}.  
An alternative spectroscopy, using planar tunnel junctions, has also made progress in circumventing charging issues \cite{Eisenstein2016, Jang2016, Zhang2017, Jang2017}. However, while an in-plane magnetic field can be used to provide momentum resolution, it is unclear whether these spectral functions  encode any analog of the discrete, topological, counting discussed here.

	Finally, another potential concern is the strength of the screening of the interactions within the 2D sample, critical to forming the FQH state, by electrons in the tip.
We believe this is unlikely; because the radius of curvature of the tip is small or comparable to $\ell_B$,  the tip will be inefficient at screening the in-plane interactions within the sample.
Thus the tip will not significantly perturb the ground state beyond the previously discussed band-bending.
However, the tip may screen the hole left behind after a tunneling event, which could be useful to model in future work.

In conclusion, we have shown that the structure of energy levels in the LDOS near an impurity encodes a sequence of integers which identifies the underlying FQH phase.
When anyonic particles are localized on the impurity, the sequence changes in a unique way, making  the LDOS  a real-space ``image'' of the location of anyons.
This new technique to study anyons can provide a new tool for exploring their use in topological quantum computing.
For instance, it can be used to detect the presence of neutral anyons, which is necessary for measuring the outcome of non-abelian braiding operations. 
While using STM spectroscopy to measure the LDOS in the FQH regime remains an open challenge, recent advances in the fabrication of graphene on boron-nitride heterostructures, and the successful application of STM to the IQHE, suggests this is a challenge  worth undertaking.

	Our work also raises a number of theoretical questions.
For certain phases, such as the recently proposed PH-Pfaffian \cite{Son2015}, we do not know how to compute the counting of the LDOS, and we hope this work will motivate future calculations.
Furthermore, while the general mathematical structure of braiding and statistics in anyon phases is well understood, how this structure relates to fractional exclusion statistics--probed here through the dimension of the several-anyon Hilbert space--is not.
The relation between exchange statistics and exclusion statistics was understood for the simplest quantum Hall phases two decades ago, but deserves renewed attention in light of our recent understanding of the interplay of symmetry and topological order \cite{barkeshli2014}.
A full understanding of fractional exclusion statistics would prove invaluable not just for predicting the LDOS, but for spectroscopic probes of topological order more generally.

\emph{Acknowledgements.}
We thank B.A.\ Bernevig, F.D.M.\ Haldane, C. Laumann, and  A.F.\ Young for helpful discussions. MPZ and AY also acknowledge the hospitality of the Aspen Center for Physics, which is supported by National Science Foundation grant PHY-1607611. AY acknowledge support from Gordon and Betty Moore Foundation as part of EPiQS initiative (GBMF4530), NSF-MRSEC programs through the Princeton Center for Complex Materials DMR-142054, and NSF-DMR-1608848. ZP acknowledges support by EPSRC grant EP/P009409/1. Statement of compliance with EPSRC
policy framework on research data: This publication is theoretical work that does not require supporting research data.


\bibliography{fqhstm}

\clearpage

\onecolumngrid
\begin{center}
\textbf{\large Supplemental Online Material for ``Imaging anyons with scanning tunneling microscopy'' }\\[5pt]
\begin{quote}
{\small In this supplementary material we present details on the exact diagonalization numerics, composite fermion calculations of the unoccupied spectral function, quantitative details on the distribution of spectral weight $w_{m, a}$, and countings for the Pfaffian phase in the presence of pinned anyons.}\\[20pt]
\end{quote}
\end{center}
\twocolumngrid

\setcounter{section}{0}

\section{Numerical methods}

The calculation of LDOS requires the exact ground state of an interacting 2DEG in the presence of a charged impurity, as well as the complete manifold of $q$-quasihole (or quasielectron) states (e.g., $q=3,4$ for Laughlin and Moore-Read states, respectively). We obtain these states using exact diagonalization in the spherical geometry~\cite{Haldane83}. This geometry conveniently allows to resolve the bulk LDOS as a function of $z$-projection of angular momentum, while avoiding edge effects. For the Laughlin and Moore-Read states, the finite-size effects are found to be weak, in particular at small values of angular momentum $m$ which contain universal topological information. The strength of the finite-size effect can be further reduced in the framework of perturbation theory~\cite{Prodan09} where the problem is projected to the manifold of the model $q$-quasihole states, which are obtained using the recursion relation of the Jack polynomials~\cite{BernevigHaldane2008}. In the Supplementary Materials  we use this method to extend the calculation by a few system sizes beyond exact diagonalization, but only allows access to the LDOS on the hole side.\footnoterecall{fn}

At filling factor $\nu=\frac{5}{2}$, PH symmetry interferes with our ability to confirm the hole counting of the Pfaffian due to the known ``aliasing" problem:
when numerically computing the states $\ket{a}$ which contribute to the LDOS, the quasihole excitations of the Pfaffian occur at the same $(N_e,N_\phi)$ as  the quasiparticle excitations of the anti-Pfaffian.
To avoid this, we have added a weak 3-body interaction to the Coulomb interaction of Eq.~\eqref{eq:vr} which breaks the PH symmetry in favor of the Moore-Read state (such perturbations, though possibly of opposite sign, are  experimentally present due to LL mixing  and disorder).
The predicted counting is confirmed in Fig.~\ref{fig:LDOS}(c).

\section{Laughlin state counting (electrons and holes) from composite fermion construction}
\label{sec:CFcounting}
\newcommand{\ltop}{{l_\textrm{top}}}
\newcommand{\LCF}{{L_\textrm{CF}}}
\newcommand{\LJa}{{L_\textrm{Jastrow}}}

It is possible to compute the excitation spectrum of the $\nu=\frac13$ Laughlin state via the composite fermion (CF) picture \cite{Jain89, Jain2005}.
The CF construction is useful in that it allows one to predict the spectrum for both holes \emph{and} electrons.

\begin{figure}[htb]
	\includegraphics[width=83mm]{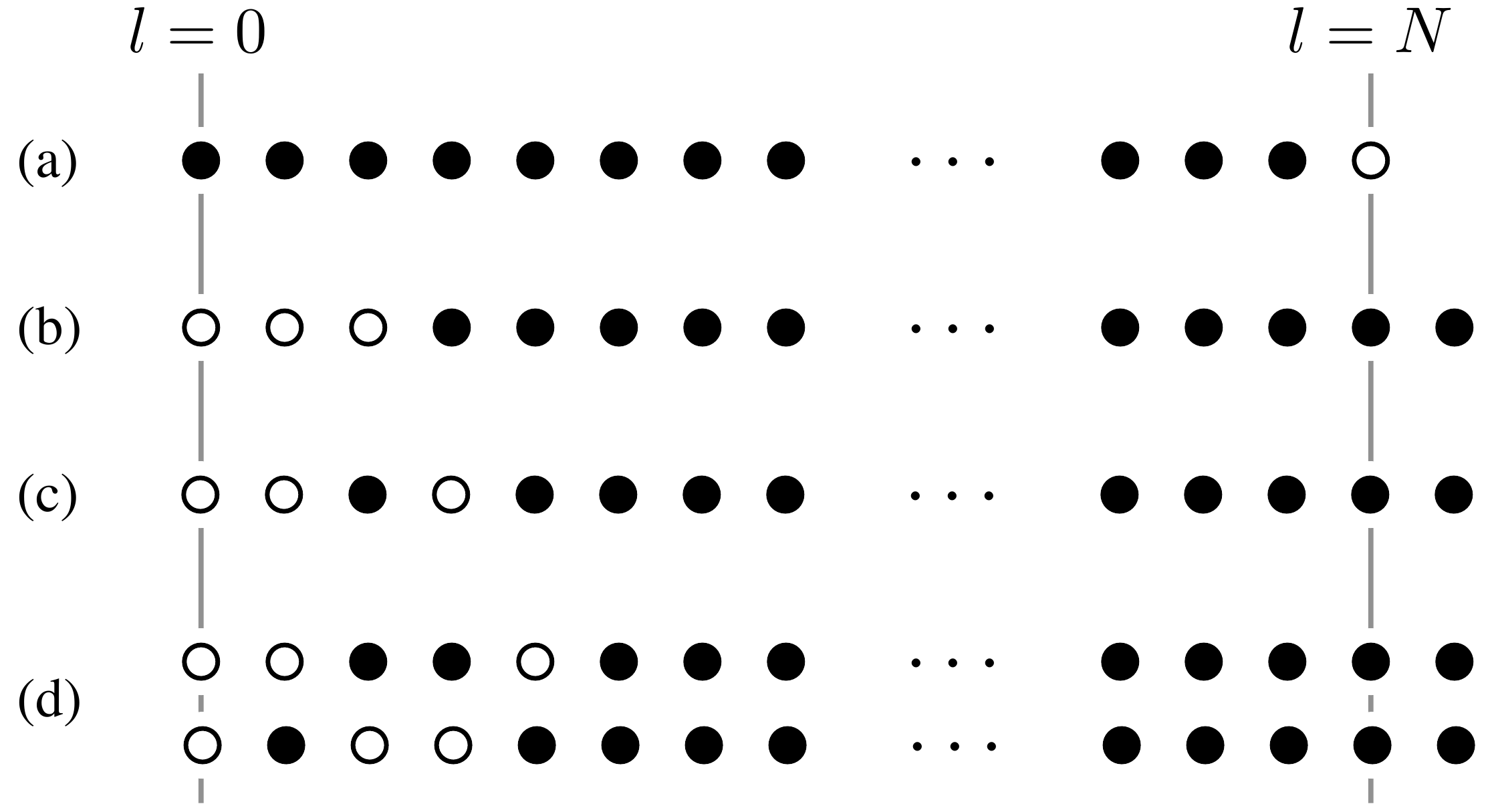}
	\caption{%
		The composite ferrmion (CF) Laughlin state with hole excitations.
		(a) The CF ground state with $N$ particles.  The CFs form an IQHE liquid occuping states $n=0$ to $n=N-1$.
		(b-d) CF states with one missing physical electron.
		The missing electron manifest itself as a missing CF along with a displacement for all the remaining CFs.
		(b) The state with $m = 0$; (c) the state with $m = 1$; and (d) the two states with $m = 2$.
	}
	\label{fig:CF:h}
\end{figure}

A CF is comprised of an electron with two flux quanta.
The Laughlin state is formed when the CFs fall into an $\nu=1$ integer quantum Hall (IQH) state.
Shown in Fig.~\ref{fig:CF:h}(a) is a CF IQH droplet, with CF orbitals $l=0,1,\ldots,N-1$ occupied.
The Laughlin wavefunction results after attaching two flux quanta to every particle, i.e., multiplying by Jastrow factor $\prod_{a<b}(z_a-z_b)^2$.
The size of the Laughlin droplet is determined by the `top angular momentum' $\ltop$, the furthest orbital occupied by any single (physical) particle.
In this case,
\begin{align}
	\ltop = (N-1) + 2(N-1) = 3N-3,
\end{align}
where $(N-1)$ is the location of the CF particle with the largest $l$, and $2(N-1)$ results from flux insertion given $N$ total particles.
The total angular momentum of this ground state is $L_0 = \LCF + \LJa$,
	where $\LCF = 0 + 1 + \ldots + (N-1) = \frac{N(N-1)}{2}$ is the total angular momentum in the CF picture,
	$\LJa = N(N-1)$ is the additional angular momentum from the flux insertion.
Thus,
\begin{align}
	L_0 = \frac{3}{2}N^2 - \frac{3}{2}N
\end{align}
for our Laughlin droplet.

\subsection{Hole excitations}

Figure~\ref{fig:CF:h}(b) shows the CF structure when one physical electron is removed, where there are now $N-1$ CFs.
Notice that each CF has been shifted by two orbitals, such that $l=3,\ldots,N+1$ are now occupied.
This is crucial to ensure that the size of the Laughlin droplet remains unchanged, that is, $\ltop = (N+1) + 2(N-2) = 3N-3$ is invariant.
We can also compute the total angular momentum $L$, and we find that $\LCF = 3 + 4 + \ldots + (N+1)$ and $\LJa = (N-1)(N-2)$.
Putting it together,
\begin{align}
	L = \LCF + \LJa = \frac{3}{2}N^2 - \frac{3}{2}N ,
\end{align}
we see that this excited state has the same angular momentum as the ground state, and hence a $m=0$ hole excitation.

Additional hole states can be constructed by moving the CFs to different orbitals, and the total angular momentum will change by that of the CFs.
For example, Fig.~\ref{fig:CF:h}(c) shows a $m=1$ ($L = L_0-1$) state, while Fig.~\ref{fig:CF:h}(d) shows the two possible $m=2$ states.
In general, each arrangement of the three empty CF orbitals will yield an excited state for the Laughlin state (with a missing electron).
The number of states at total momentum $L = L_0 - m$ is given by the number of ways to decompose $m+3$ as a sum of 3 distinct integers,
	or equivalently, number of ways to decompose $m$ as a sum of 3 integers.
This method yields the sequence $1,1,2,3,4,5,\dots$, in agreement with the $(1,3)$-exclusion rule.

\subsection{Electron excitations}
\begin{figure}[t]
	\includegraphics[width=83mm]{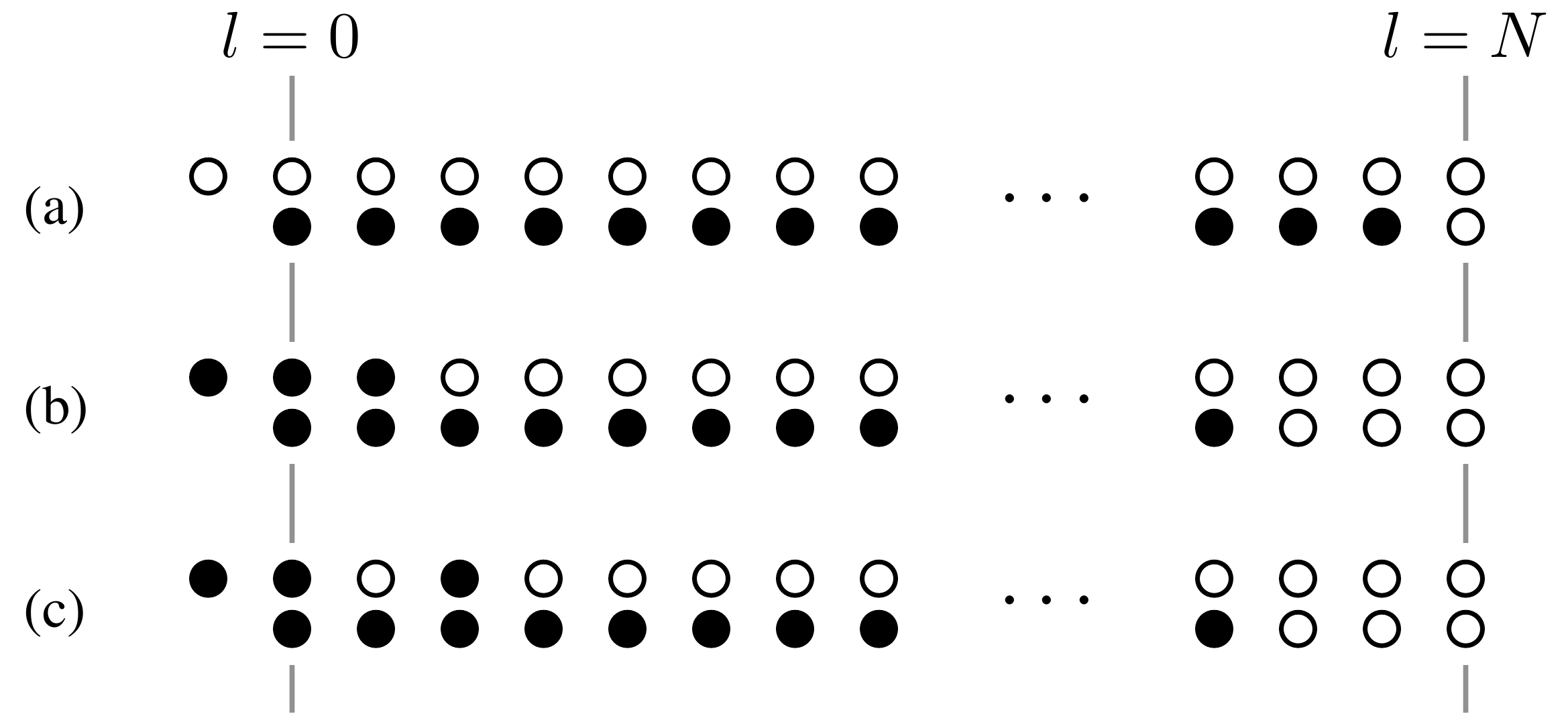}
	\caption{%
		The composite ferrmion (CF) Laughlin state with electron excitations.
		(a) The CF ground state with $N$ particles.  The CFs are all occupying the lowest CF Landau level.
		(b-c) CF states with one additional physical electron.
		Now there are $N+1$ CFs, and three of them are in an excited CF Landau level.
		(b) The lowest excited state with $m = 3$; and (c) with $m = 4$.
	}
	\label{fig:CF:e}
\end{figure}

When an electron is inserted in the Laughlin droplet, the CFs will need to occupy more than the lowest CF Landau level; some CFs will be placed into the first excited CF Landau level.
Placing CFs in multiple Landau levels is part of the standard construction for hierarchy states, e.g.\ two filled CF Landau levels yields the $\nu = \frac25$ hierarchy state.
The resulting wavefunction is the CF wavefunction, multiplied by the Jastrow factor, and projected into the lowest Landau level \cite{Jain89}.

Figure~\ref{fig:CF:e}(a) again shows a Laughlin droplet (ground state) with $N$ particles.
This figure is the same as Fig.~\ref{fig:CF:h}(a), with the unoccupied excited CF Landau level also shown.

When an electron is added to the ground state, there are now $N+1$ CFs in the system.
The orbitals in the lowest CF Landau level $l=0,\ldots,N-3$ are occupied.
It is straightforward see that $\ltop$ of this state remains the same (that is, $3N-3$), and no further CFs can be added to the level, lest the size of the Laughlin droplet be increased.
Hence, there must be at least three CFs in an excited Landau level.
If we are to interpret the CF Landau levels as energies levels, the lowest excitations would consists of exactly three CFs in the 1\textsuperscript{st} excited Landau level.
Figure~\ref{fig:CF:e}(b) and (c) are examples of such excited states.

\begin{figure}[t]
	\includegraphics[width=65mm]{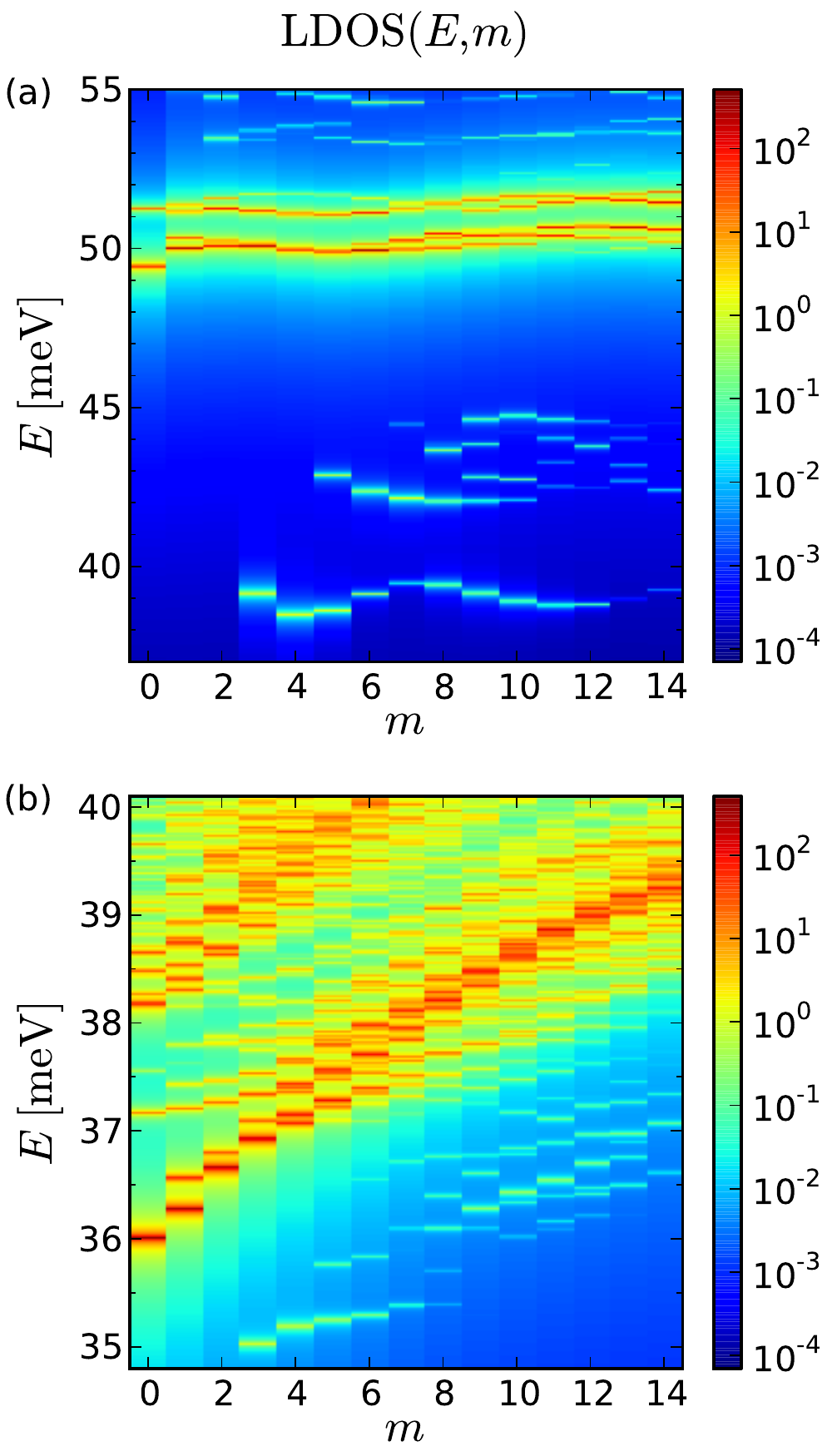}
	\caption{
		LDOS on the electron side of the Laughlin $\nu=\frac{1}{3}$ state.
		(a) Pseudopotential interaction with $V_1=1$ and a delta interaction impurity of strength 0.2 above the north pole of the sphere.
		(b) Screened Coulomb interaction with gate at a distance $d_g=4\ell_B$ and impurity at $d_i=2.5\ell_B$.
		In both cases, the system contains $N_\phi=24$ flux quanta.
		The universal counting $0,0,0,1,1,2,\cdots$, predicted by CF theory, is observed in the lowest band which, however, carries much less spectral weight than the band above it.
		The band which dominates the spectral weight is expected to merge into a continuum in the thermodynamic limit. 
	}
	\label{fig:LDOSm_Laughlin_e}
\end{figure}

The CF configurations in the Fig.~\ref{fig:CF:e}(b) have the least total $L$, and is clearly the unique configuration with such angular momentum.
For this state, $\LCF = (-1+0+1) + [0+\ldots+(N-3)] = \frac{(N-3)(N-2)}{2}$, while the Jastrow factor contributes $\LJa = N(N+1)$, for a total
\begin{align}\begin{split}
	L &= \LCF + \LJa = \frac{3}{2}N^2 - \frac{3}{2}N + 3
	\\	&= L_0 + 3 .
\end{split}\end{align}
Hence for the lowest electron-excited states of the Laughlin state has relative angular momentum $m=3$, and all other states with larger $m$.
Once again, we can map the counting of states to partition of $m$ into three integers, and therefore the counting (starting at $m=0$) is $0,0,0,1,1,2,3,\dots$, with \emph{no low energy states} for $m \leq 2$.
Figure~\ref{fig:LDOSm_Laughlin_e} shows the electron-excited spectrum for the Laughlin state.
Seen underneath the ``main branch'' are states, starting at $m=3$, consistent with our counting.
These states all have low spectral weights, which may be attributed to the dominant coupling of the electron operator to the further-excited $m=0$ state (e.g. at $\approx \unit[49.4]{meV}$ in \ref{fig:LDOSm_Laughlin_e}a).

\section{Distribution of spectral weight with varying impurity distance}

The tunneling current into level $a$ is proportional to the spectral weight $w_{a, m} = |\!\braket{a | \hat{c}_m | \Omega}\!|^2$, which satisfies the sum rule $\sum_a w_{a, m} = \langle \hat{n}_m \rangle \approx \nu$.
The distribution of spectral weight depends on the details of the interactions and impurity potential, which may require optimization for maximum visibility: ideally the weight would be equally distributed across all the levels.
The ideal situation is in fact nearly realized for  short-range pseudopotential  Hamiltonians~\cite{Haldane83}, but for Coulomb interactions we find an uneven distribution of weight.
As can be seen from Fig.~\ref{fig:LDOS}(b), the weight is greatest for the high energy level, which roughly consists of three quasiholes  directly on top of each other,  simultaneously maximizing both their Coulomb repulsion and the overlap with the electron operator $\psi(\mathbf{r})$.
We find that  the spectral weight is the most unevenly distributed when the interactions are very strong compared to the impurity (e.g., $\dg, \di \gg \ell_B$); indeed, in the absence of the impurity, only the dominant high-energy level has weight and the multiplicity is undetectable.
On the other hand, if the impurity becomes too strong compared to the interactions, it will become favorable for a quasiparticle to bind to the impurity, changing the counting. 
We find, for instance, that $w_{3, a}$ at best ranges over two orders of magnitude.
Because of the increasingly small weight, it may prove challenging to measure the counting beyond the first several $m$ sectors.
However, this does not seem to be a fundamental limitation since all the cases discussed in this paper are distinguished by the counting of only the first two orbitals.

\begin{figure}[t]
	\includegraphics[width=\linewidth]{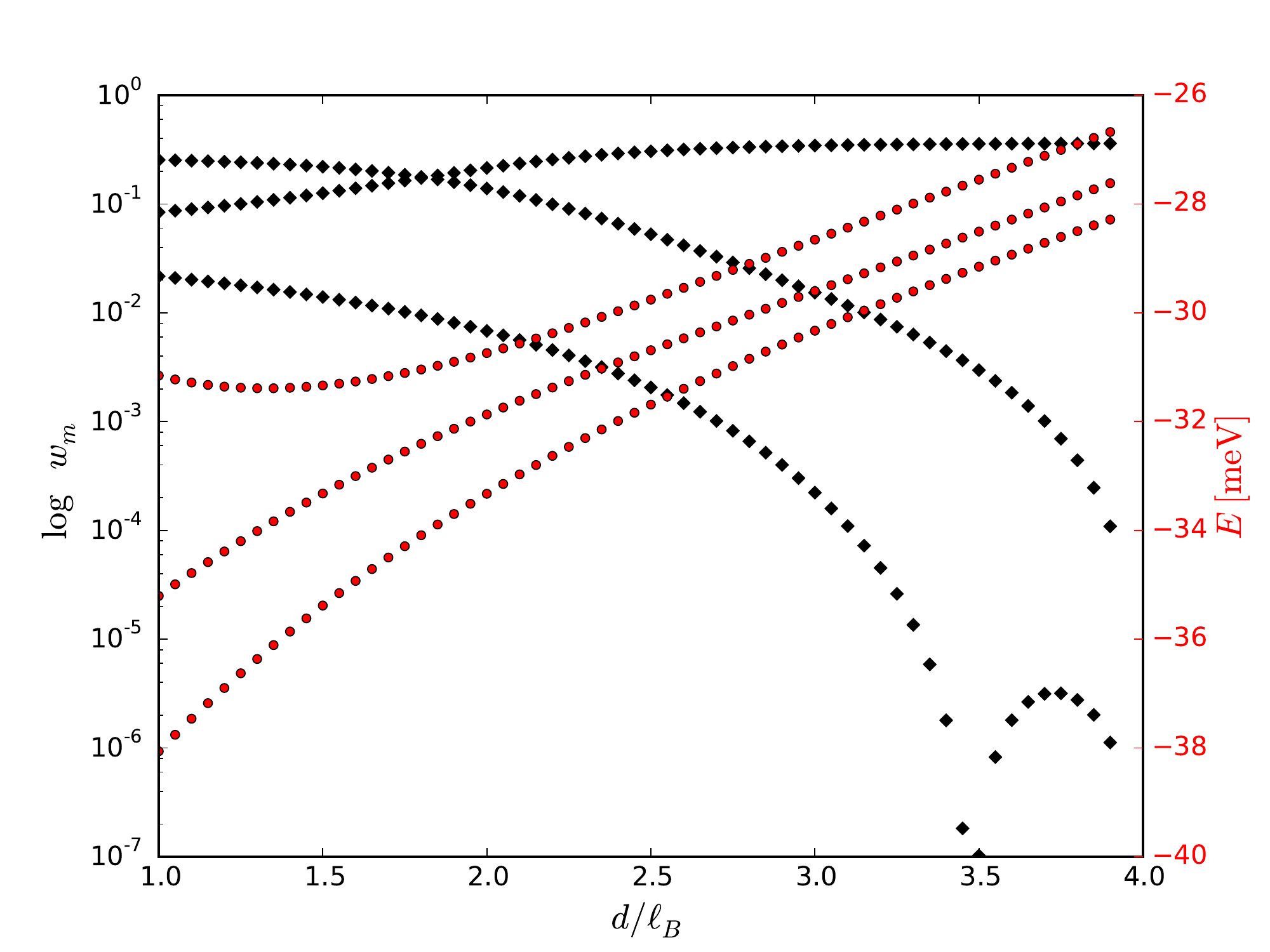}
	\caption{\label{fig:weightsm}Perturbative calculation of spectral weight $w_m$ and energy for the three states belonging to a sector $m=3$, as a function of impurity distance $d$.
		The calculation is performed by projecting the Coulomb interaction (with gate at $d_g=4\ell_B$ and $N_\phi=24$) onto the manifold of exact quasihole states with $m=3$ (e.g., obtained as zero-energy eigenstates of the $V_1$ pseudopotential Hamiltonian), and diagonalizing the corresponding $3\times 3$ matrix.
		We observe that tuning $d$ has an opposite effect on energy vs.\ spectral weight: for small $d$, the degeneracy of the three states is strongly lifted, while at large $d$ the spectral weight is very unevenly distributed between the three states.
		We identify the impurity distance $d\sim 2.5\mbox{--}3\ell_B$ as the optimal one for measuring LDOS.
		Similar conclusion can be drawn by studying other $m$ sectors, or performing a full (non-perturbative) calculation. }
\end{figure}

	Here we provide additional quantitative information on the low-energy spectral weights $w_{a, m} = |\!\braket{a | \hat{c}_m | \Omega}\!|^2$.
Fig.~\ref{fig:weightsm} presents an approximate calculation of the $w_{3, a}$ weights as a function of the impurity distance $\di$; we indeed see that weight is more evenly distributed for stronger impurities.
The choice  of $\di = 2.7\ell_B$, $\dg = 4\ell_B$ used in Fig.~\ref{fig:LDOS}(b) of the main text was found by fixing $\dg$ and tuning through $\di$ in order to find a value where the weights were most uniformly distributed without binding a quasiparticle~\footnoterecall{fn}.
At this value  the $\nu=\frac{1}{3}$ weights in the $m=2$ orbital are $w_{2, a} = \{ 0.34, 0.023 \}$, and in the $m=3$ orbital, $w_{3, a} = \{0.32, 0.06, 0.003\}$.
The spectral weights for higher $m$ have larger distributions, leading to small tunneling currents which would be difficult to measure. 
Fortunately all the cases we have considered are distinguished by the counting of even just $w_0$ and $w_1$.

\section{Countings of Pfaffian phase in presence of pinned anyons}
\label{sec:excl_counting}

The Pfaffian (Moore-Read) phase satisfies the $(2,4)$-rule, with the root configuration of the ground state given by $11001100\cdots$.
Table~\ref{tab:MRcounting} shows the counting of occupied states in the presence of various pinned quasiparticles.
\begin{table}[b]
	\begin{tabular}{c|l|l}
		\begin{tabular}{c}Trapped\\[-0.5ex]quasiparticle\end{tabular} & \begin{tabular}{c}Root\\[-0.5ex]Configuration\end{tabular} & Counting
	\\\hline
		None & $1100110011\cdots$ & 1,2,3,5,7,9,14,\dots
	\\	Neutral fermion & $1011001100\cdots$ & 1,1,3,4,6,8,13,\dots
	\\	$\frac{e}{4}$ anyon & $1010101010\cdots$ & 1,2,3,6,8,15,20,\dots
	\\	$-\frac{e}{4}$ anyon & $1101010101\cdots$ & 1,2,2,4,5,6,11,\dots
	\\	$\frac{e}{2}$ (flux) & $0110011001\cdots$ & 0,1,2,4,6,10,14,\dots
	\\	$\frac{e}{2}$ (3 flux + electron) & $1001100110\cdots$ & 1,1,3,5,8,12,20,\dots
	\end{tabular}
	\caption{Counting (hole-side) of the Moore-Read state.}
	\label{tab:MRcounting}
\end{table}

\clearpage	

\end{document}